\documentclass[lettersize,journal]{IEEEtran}
\usepackage{amsmath,amsfonts}
\usepackage{algorithmic}
\usepackage{algorithm}
\usepackage{array}
\usepackage[caption=false,font=normalsize,labelfont=sf,textfont=sf]{subfig}
\usepackage{textcomp}
\usepackage{stfloats}
\usepackage{url}
\usepackage{verbatim}
\usepackage{listings}
\usepackage{graphicx}
\usepackage{multirow}
\usepackage{rotating}
\usepackage{booktabs}

\usepackage[utf8]{inputenc}
\usepackage{amsmath}
\usepackage{float}
\usepackage{amsfonts}
\usepackage{amssymb}
\usepackage{soul}
\usepackage{xcolor}

\usepackage[
	backend=biber, % Use the biber backend for compiling the bibliography
	citestyle=numeric, % In-text citation style
	bibstyle=numeric, % Bibliography style
	sorting=none, % Order references in the order in which they are used in the document
]{biblatex}
\begin{document}

\title{AcceleratedKernels.jl: Cross-Architecture Parallel \\ Algorithms from a Unified, Transpiled Codebase}

\author{Andrei-Leonard Nicușan\textsuperscript{1}, Dominik Werner\textsuperscript{1}, Simon Branford\textsuperscript{2}, Simon Hartley\textsuperscript{2}, \\ Andrew J. Morris\textsuperscript{3}, Kit Windows-Yule\textsuperscript{1} \\[1em]

\footnotesize\textsuperscript{\textbf{1}}School of Chemical Engineering, University of Birmingham, UK

\footnotesize\textsuperscript{\textbf{2}}Advanced Research Computing, University of Birmingham, UK

\footnotesize\textsuperscript{\textbf{3}}School of Metallurgy and Materials, University of Birmingham, UK

\thanks{}
\thanks{Manuscript received on \today.}}

\markboth{Manuscript - \today}%
{}

\IEEEpubid{}

\maketitle

\begin{abstract}

AcceleratedKernels.jl is introduced as a backend-agnostic library for parallel computing in Julia, natively targeting NVIDIA, AMD, Intel, and Apple accelerators via a unique transpilation architecture. Written in a unified, compact codebase, it enables productive parallel programming with minimised implementation and usage complexities. Benchmarks of arithmetic-heavy kernels show performance on par with C and OpenMP-multithreaded CPU implementations, with Julia sometimes offering more consistent and predictable numerical performance than conventional C compilers. Exceptional composability is highlighted as simultaneous CPU-GPU co-processing is achievable - such as CPU-GPU co-sorting - with transparent use of hardware-specialised MPI implementations. Tests on the Baskerville Tier 2 UK HPC cluster achieved world-class sorting throughputs of 538-855 GB/s using 200 NVIDIA A100 GPUs, comparable to the highest literature-reported figure of 900 GB/s achieved on 262,144 CPU cores. The use of direct NVLink GPU-to-GPU interconnects resulted in a 4.93x speedup on average; normalised by a combined capital, running and environmental cost, communication-heavy HPC tasks only become economically viable on GPUs if GPUDirect interconnects are employed.

\end{abstract}


\section{Introduction}

The development of high-performance parallel algorithms across heterogeneous computing architectures poses significant challenges in scientific computing \cite{kachris2016survey}. In this context, a kernel is the core piece of code that runs a key operation such as matrix multiplication, sorting, or image processing, often being executed across many elements repeatedly and/or concurrently. Traditional methods often require platform-specific code or rely heavily on vendor-supported libraries, leading to increased complexity and maintenance burdens for both implementors and users. In real-world applications, users often work with heterogeneous hardware, ranging from multithreaded shared-memory workstations to distributed data centre GPUs. Writing platform-specific code for each environment increases complexity and reduces the each platform's adoption, in turn minimising their possible performance benefits and siloing developed codes.

AcceleratedKernels.jl addresses these challenges by providing a novel backend-agnostic, cross-architecture library for parallel computing within the high-level, high-performance Julia programming language, being the first high-productivity approach that natively targets NVIDIA, AMD, Intel and Apple accelerator hardware via transpilation. This ensures portability while maintaining performance, allowing users to maximise the capabilities of their available hardware, independent of specific vendor ecosystems. AcceleratedKernels.jl has been adopted by the official JuliaGPU organisation; it is available as an open-source library at https://github.com/JuliaGPU/AcceleratedKernels.jl.

\begin{table*}[htbp]
  \centering
  \caption{Comparison of cross-architecture programming models in active use}
    \begin{tabular}{llrrrrrrrr}
          &       &       & \multicolumn{4}{c}{\textbf{GPU Hardware Supported}} & \multicolumn{1}{p{3.25em}}{\textbf{Intrinsics}} & \multicolumn{2}{l}{\textbf{Burden / Complexity}} \\
    \textbf{Type} & \textbf{Framework} & \multicolumn{1}{l}{\textbf{Usage}} & \multicolumn{1}{l}{Nvidia} & \multicolumn{1}{l}{AMD} & \multicolumn{1}{l}{Intel} & \multicolumn{1}{l}{Apple} & \multicolumn{1}{p{3.25em}}{\textbf{Access}} & \multicolumn{1}{l}{Implementor} & \multicolumn{1}{l}{User} \\
    \midrule
    \multirow{5}[2]{*}{Standard} & OpenCL & \multicolumn{1}{p{12.585em}}{Separate-source kernels} & \multicolumn{1}{l}{Yes} & \multicolumn{1}{l}{Yes} & \multicolumn{1}{l}{Yes} & \multicolumn{1}{l}{No***} & \multicolumn{1}{l}{Yes} & \multicolumn{1}{l}{High} & \multicolumn{1}{l}{High} \\
          & OpenMP & \multicolumn{1}{p{12.585em}}{Commented directives} & \multicolumn{1}{l}{Yes} & \multicolumn{1}{l}{Yes} & \multicolumn{1}{l}{Yes} & \multicolumn{1}{l}{No} & \multicolumn{1}{l}{No} & \multicolumn{1}{l}{High} & \multicolumn{1}{l}{Low} \\
          & OpenACC & \multicolumn{1}{p{12.585em}}{Commented directives} & \multicolumn{1}{l}{Yes} & \multicolumn{1}{l}{Yes} & \multicolumn{1}{l}{No} & \multicolumn{1}{l}{No} & \multicolumn{1}{l}{No} & \multicolumn{1}{l}{High} & \multicolumn{1}{l}{Low} \\
          & Vulkan & \multicolumn{1}{p{12.585em}}{Separate-source kernels} & \multicolumn{1}{l}{Yes} & \multicolumn{1}{l}{Yes} & \multicolumn{1}{l}{Yes} & \multicolumn{1}{l}{Yes} & \multicolumn{1}{l}{Yes} & \multicolumn{1}{l}{High} & \multicolumn{1}{l}{High} \\
          & SYCL  & \multicolumn{1}{p{12.585em}}{Single-source kernels} & \multicolumn{1}{l}{Yes****} & \multicolumn{1}{l}{Yes****} & \multicolumn{1}{l}{Yes****} & \multicolumn{1}{l}{No} & \multicolumn{1}{l}{Yes} & \multicolumn{1}{l}{High} & \multicolumn{1}{l}{Medium} \\
    \midrule
    \multirow{3}[2]{*}{API} & Kokkos & \multicolumn{1}{p{12.585em}}{Library functions and C++ lambda simple loops} & \multicolumn{1}{l}{Yes} & \multicolumn{1}{l}{Yes} & \multicolumn{1}{l}{Yes*} & \multicolumn{1}{l}{No} & \multicolumn{1}{l}{No} & \multicolumn{1}{l}{Medium} & \multicolumn{1}{l}{Medium} \\
          & RAJA  & \multicolumn{1}{p{12.585em}}{Library functions and C++ lambda simple loops} & \multicolumn{1}{l}{Yes} & \multicolumn{1}{l}{Yes} & \multicolumn{1}{l}{Yes*} & \multicolumn{1}{l}{No} & \multicolumn{1}{l}{No} & \multicolumn{1}{l}{Medium} & \multicolumn{1}{l}{Medium} \\
          & ArrayFire & \multicolumn{1}{p{12.585em}}{Library functions and JIT-compiled simple loops} & \multicolumn{1}{l}{Yes} & \multicolumn{1}{l}{Yes**} & \multicolumn{1}{l}{Yes} & \multicolumn{1}{l}{No***} & \multicolumn{1}{l}{No} & \multicolumn{1}{l}{Medium} & \multicolumn{1}{l}{Low} \\
    \midrule
    \multicolumn{1}{l}{\multirow{3}[2]{*}{Language}} & Halide & \multicolumn{1}{p{12.585em}}{Functional C++ DSL for image processing kernels} & \multicolumn{1}{l}{Yes} & \multicolumn{1}{l}{Yes} & \multicolumn{1}{l}{Yes} & \multicolumn{1}{l}{Yes} & \multicolumn{1}{l}{No} & \multicolumn{1}{l}{Medium} & \multicolumn{1}{l}{Medium} \\
          & Futhark & \multicolumn{1}{p{12.585em}}{Functional language for simple MapReduce-like kernels} & \multicolumn{1}{l}{Yes} & \multicolumn{1}{l}{Yes**} & \multicolumn{1}{l}{Yes**} & \multicolumn{1}{l}{No***} & \multicolumn{1}{l}{No} & \multicolumn{1}{l}{Medium} & \multicolumn{1}{l}{Medium} \\
          & Bend/HVM2 & \multicolumn{1}{p{12.585em}}{Combinator-based functional language} & \multicolumn{1}{l}{Yes} & \multicolumn{1}{l}{No} & \multicolumn{1}{l}{No} & \multicolumn{1}{l}{No} & \multicolumn{1}{l}{No} & \multicolumn{1}{l}{Medium} & \multicolumn{1}{l}{Low} \\
    \midrule
    Transpiler & \multicolumn{1}{p{9.415em}}{AcceleratedKernels.jl / KernelAbstractions.jl} & \multicolumn{1}{p{12.585em}}{Library functions and high level single-source kernels} & \multicolumn{1}{l}{Yes} & \multicolumn{1}{l}{Yes} & \multicolumn{1}{l}{Yes} & \multicolumn{1}{l}{Yes} & \multicolumn{1}{l}{No} & \multicolumn{1}{l}{Low} & \multicolumn{1}{l}{Low} \\
\cmidrule{1-2}    * via SYCL & *** deprecated &       &       &       &       &       &       &       &  \\
    ** via OpenCL & **** Linux only &       &       &       &       &       &       &       &  \\
    \end{tabular}%
  \label{tab:models}%
\end{table*}%

\subsection{Alternative Approaches}
\label{sec:alternatives}

As summarised in Table \ref{tab:models}, there are currently three main approaches to cross-architecture algorithm development used in production, or mainstream code today: i) \textbf{standards-based}, which define an abstract framework - as a set of APIs, libraries and/or compilers - which is then left to hardware and software developers to implement, ii) \textbf{API-based}, wherein a unified library interface is defined, which abstracts calls to different existing libraries for individual backends, and iii) \textbf{programming language-based}, where a domain-specific language (DSL), or an esoteric programming language is created for kernel-writing, which are then compiled directly, or transpiled to an existing software stack. Each approach has different trade-offs, most starkly in the implementation and usage complexities, which in turn affect resulting code quality and performance, and adoption.

For example, while the popular OpenMP and OpenACC frameworks are some of the most accessible approaches available today - owing to their non-invasive usage, wherein standard C++ code is annotated with compiler directives, and stack maturity within standard compilers (OpenMP is available within the default Clang and GCC toolchains) - they require very high implementation and maintenance efforts, indeed necessitating continuous updating as the general-purpose languages evolve; however, in the end, they are also perhaps the least flexible approaches, being limited to relatively simple looping constructs with possible reduction operators \cite{dagum1998openmp, wienke2012openacc, antao2016offloading, novillo2006openmp}. The other popular frameworks in this category in active use are the OpenCL, SYCL and Vulkan standards, currently developed by the Khronos consortium, which define C/C++ dialects for kernel-writing; a main advantage over the previously-mentioned approaches is the algorithm writing flexibility, as they map fairly closely to native GPU constructs. However, they are wholly-reliant on good implementations from industry, each typically requiring direct, heavy involvement from the hardware manufacturer; for example, OpenCL kernels have long been resulting in lower performance on Nvidia hardware than the native CUDA kernel language, though at present their performance is similar \cite{komatsu2010evaluating, holm2020performance}. Moreover, disagreements over the standard evolution can result in deprecation on entire platforms, such as OpenCL on Apple devices; SYCL is not currently supported in production-level compiler toolchains on Apple Silicon chips - similarly, the only mature SYCL implementation is the Intel DPC++ / oneAPI offering, currently only available on Linux machines \cite{reinders2021data}. The only standard widely available on all architectures and operating systems considered in this study is Vulkan, which is a successor to the famous OpenGL rendering API, while being focused on extremely fine control over hardware, and includes GPGPU compute capabilities - as such, it is also by far the most difficult to use, typically requiring hundreds of lines of repetitive, ``boilerplate'' code for setting up devices, contexts, command queues, etc.; there is currently no production-level scientific code using Vulkan as its compute backend \cite{sellers2016vulkan}. Finally, it also requires that Vulkan code be written as separate sources - either files or inline strings - which is an often-cited reason for OpenCL's difficult adoption in the scientific space \cite{rupp2016opencl}.

Compared to standards-based approaches, unified APIs are most often implemented by third parties, without requiring direct support from hardware manufacturers. Instead, they redirect function calls from a uniform interface to specific libraries for each backend; these backend libraries, though, are often officially-endorsed by the hardware manufacturers (e.g. Kokkos uses the Nvidia Thrust parallel primitives library for the CUDA backend, and AMD rocThrust for the ROCm backend) - thus, while the implementation effort is lower, the maintenance burden increases, as the API must reactively track changes in the backend libraries. The two most actively-used ``programming models'' of the API approaches are Kokkos and RAJA, developed by the Sandia and Lawrence-Livermore US National Labs - thus showing that high investment is still required in this space \cite{edwards2014kokkos, beckingsale2019raja}. ArrayFire, another popular API, especially through its wrappers in higher-level scripting languages like Python, is also backed by AccelerEyes LLC, a private company \cite{malcolm2012arrayfire}. The main advantage of API approaches is the excellent performance that results from using the official parallel primitives for each backend, while also being easier to use than standards like OpenCL or Vulkan; however, their flexibility is only slightly better than directives-based frameworks such as OpenMP or OpenACC, as only the ``greatest common denominator'' of the algorithms offered by the backend libraries can be exposed. For example, while the ``\texttt{upper\_bound}'' function for finding the insertion indices of some elements in a sorted vector, while maintaining ordering (called ``\texttt{searchsorted}'' in other programming languages), is implemented within the Nvidia Thrust library, it is not included in any of the API-based programming models, as not all backend libraries include it - though it is required, for example, in the ``MPISort'' algorithm benchmarked in Section \ref{sec:mpisort}; others, like ``\texttt{sortperm}'' for getting an index permutation which sorts an array, are simply not available, and implementing them on top of the default sorting interface requires either a custom comparator (e.g. unavailable in Kokkos) or unnecessary data copies. Besides these functions, all APIs offer the possibility of coding simple kernels as annotated C++ anonymous functions (lambdas), or functors (classes with overloaded `\texttt{()}` operators) which fall under the ``\texttt{foreach}'', ``\texttt{reduce}'', and ``\texttt{scan}'' algorithm categories. Finally, Kokkos and RAJA also focus on distributed memory - across MPI ranks on supercomputing clusters - accessed via a uniform interface, which simplifies the implementation of large-scale simulation codes (typically mesh-based, like CFD and FEA), a prime example of which being the Trilinos suite of numerical algorithms, built on Kokkos \cite{heroux2012new}.

Domain-specific languages (DSLs) and esoteric programming languages are also interesting approaches, with many important compiler advances over the years first appearing in research-driven esoteric compilers; among them, three fairly recent ones are noteworthy for currently seeing some adoption: Halide, Futhark and Bend. While being more pleasant to use than the other lower-level approaches built on / around C++, they all use immutable functional constructs to model computation as forms of directed acyclic graphs (DAGs) - which, on one hand allow easier parallelisation of code, but on the other, result in much higher memory use and unnecessary data copies than mutable approaches, relying on complex compilation passes to improve thereupon. In general, compute-intensive tasks as used in numerical algorithms can be much faster when written as in-memory mutating code, than immutable data transformations; these are important considerations as GPU VRAM memory is much more expensive than RAM. Finally, none of the languages above provide direct access to the hierarchical memory offered by modern GPUs - global, shared and private - which is crucial in achieving the high throughput possible on such accelerators \cite{kurzak2010scientific}.

\subsection{Significance of this Work}
\label{sec:significance}

The AcceleratedKernels.jl library is the first truly cross-architecture standard library of parallel algorithms from a unified, transpiled codebase; a unique aspect of the library, built on the KernelAbstractions.jl backend-agnostic Julia-based kernel language, is that it is \textit{transpiled} to the native intermediate representation (IR) of each platform (PTX on Nvidia devices, AIR on Apple GPUs, other LLVM IR dialects for AMD and Intel) - thus ensuring similar performance to the official toolchains \cite{besard2018effective}. Stemming from this, a number of advantages are highlighted:

\begin{itemize}
    \item Benefitting from Julia-specific optimisations, as well as all the optimisation work poured into the official compiler stacks.
    \item As Julia is a natively homoiconic programming language - exposing its own source code as data that can be manipulated, similar to Lisp - the implementation burden of running Julia code on GPUs is significantly lower than all previous approaches \cite{bezanson2017julia}.
    \item Exceptional degree of reusability in the JuliaGPU transpilation middleware, such that new backends can be added in the future as new architectures are developed (e.g. TPUs, FPGA-like devices) with much lower effort, further decreasing the implementor-side complexity (each individual backend is implemented as a relatively small Julia library, e.g. AMDGPU.jl, oneAPI.jl, Metal.jl).
    \item Very good flexibility in kernel development, as - similar to the AcceleratedKernels.jl algorithms - highly-specialised algorithms can be written in the same language, single-source.
    \item Exceptional composability with other Julia codes - for example, many functions from the Julia Base standard library can be called directly from within kernels, which are then inlined and transpiled to the selected GPU backend with no special-casing required. See Section \ref{sec:mpisort} for an example using an external MPI-based sorting library which can use Julia Base sorters concurrently with GPU sorters to achieve simultaneous CPU-GPU co-processing - again, with no special-casing on either library's side.
    \item Very good numerical performance, on par with (and, perhaps surprisingly, sometimes exceeding) C and OpenMP CPU codes, and on the same order of magnitude as official Nvidia parallel primitives libraries (see the arithmetic-heavy kernels in Section \ref{sec:arithmetic} and the multi-node/multi-device sorting benchmark in Section \ref{sec:mpisort}).
    \item Another unexpected finding is that, for numerics, Julia can offer better consistency in the performance of the compiled code than C (see the arithmetic kernels benchmark in Section \ref{sec:ljg}; code available in supplementary materials).
    \item Following Julia's on-demand compilation model - wherein code is compiled only upon use (as opposed to ahead-of-time compilation which must compile all possible usage permutations, or just-in-time compilation which starts in an interpreted mode) - code can be highly generic, with most types not requiring explicit specification, similar to everything being a C++ template argument by default; thus Julia compilation results in excellent inlining of code defined across external libraries (as opposed to per-translation unit as in traditional compilers) and optimisation for the given types, at use-time.
    \item Being implemented in a mainstream language facilitates i) its ease of use, ii) low-effort maintenance, iii) performance improvements, and iv) adoption, each of the aforementioned having a synergetic effect on the others.
\end{itemize}

In terms of `real-world' significance, sorting processes are central to a number of important applications, ranging from simply the processing of large data sets (as is increasingly common in the age of AI), to collision detection in autonomous vehicles \cite{li2024autonomous}, to the simulation of molecular/atomistic systems using molecular dynamics (MD) \cite{thompson2022lammps} or industrial and geological systems using the discrete element method (DEM) \cite{windows2016numerical}. Indeed, for many such applications, the sorting step is the most significant bottleneck in the entire computation \cite{Windows-Yule2024}. Considering the widespread use of these techniques, the significant reduction in computational load, and thus compute power consumption, facilitated by AcceleratedKernels.jl stands to carry non-trivial sustainability benefits.

The main disadvantage of the current transpiler approach is that platform-specific intrinsics are not available - for example, per-warp shuffle instructions are not exposed in the KernelAbstractions.jl Julia-based kernel language, which are useful in improving the performance of reductions; for other algorithms, such as radix sort, intrinsics are essential for high performance. Further, the sync-cooperative thread-group size of GPUs (``warp'' in Nvidia nomenclature, ``wavefront'' for AMD) is not currently exposed, which again could help in ``reduce'' and ``mapreduce'' algorithms. While they are possible future additions, the trade-off for the excellent platform support (as detailed in the previous section, the best after Vulkan) is deemed more valuable; still, even in the absence of intrinsics, which would by definition hinder cross-platform unified codebases, performance is either similar to, or on the same order of magnitude as, officially-endorsed parallel primitives.

\section{Library Architecture}
\label{sec:architecture}

AcceleratedKernels.jl is written as a collection of functions following the Julia Base naming conventions; as Julia uses multiple dispatch to ``overload'' function names based on the complete set of calling arguments and types (each specialised implementation called a method), there is no collision between the two. For example, while ``\texttt{mapreduce(f, op, itr)}'' is defined within Base Julia - note that the types of the unary function ``\texttt{f}'', binary reduction operator ``\texttt{op}'' and iterator ``\texttt{itr}'' are not explicitly defined, and are therefore the most generic - we can implement ``\texttt{mapreduce(f, op, itr::AbstractGPUVector})'', such that ``\texttt{itr}'' subtypes of ``\texttt{AbstractGPUVector}'' will result in the AcceleratedKernels.jl specialised function being called; thus, CPU arrays (e.g. ``\texttt{Vector}'', ``\texttt{SVector}'') will be dispatched to the Julia base method, while GPU vectors (``\texttt{ROCArray}'', ``\texttt{oneArray}'', ``\texttt{MtlArray}'', ``\texttt{CuArray}'') will result in the accelerated method being used. Note that in these cases, unlike C++ virtual methods that have runtime dispatch based on an internal ``vtable'', if the types are known at call-time, the multiple dispatch mechanism is static, completed at compile time.

\begin{algorithm}[H]
\caption{Low-level form KernelAbstractions.jl copy kernel implementation.}\label{alg:copy_la}
\begin{lstlisting}
using KernelAbstractions
@kernel function copy_ka!(dst, @Const(src))
    iblock = @index(Group,Linear)
    ithread = @index(Local,Linear)
    block_size = @groupsize()
    iglobal = ithread+(iblock-1)*block_size
    dst[iglobal] = src[iglobal]
end
\end{lstlisting}
\label{alg:copy_ka}
\end{algorithm}

The KernelAbstractions.jl Julia-based kernel language is compact, mapping closely to the constructs used in General-Purpose GPU (GPGPU) programming e.g. CUDA (OpenCL), such as threads (workitems), thread blocks (workgroups), block grids (ndrange), an example of which is given in Algorithm \ref{alg:copy_ka}; note that types do not have to be explicitly defined, and instead the method will be specialised for each individual calling set of types (e.g. \texttt{MtlArray\{Float32\}}, \texttt{ROCArray\{Int128\}}, \texttt{oneArray\{CustomStruct\}}) \cite{Churavy_KernelAbstractions_jl}. Besides the constructs for querying the thread and block indices, block and grid sizes, intra-block synchronisation, and shared and private memory allocation, almost all normal Julia code is permitted within kernels - notable exceptions being that values cannot be returned, dynamic memory allocation is not permitted outside of shared memory, and exceptions cannot be thrown.

Owing to the high-level designs of the Julia language and KernelAbstractions.jl, the former's type genericity and multiple dispatch mechanisms, the AcceleratedKernels.jl library architecture is fairly simple and compact, using normal Julia constructs and a relatively flat architecture. It can be installed using ``\texttt{Pkg}'', the built-in Julia package manager; individual backends can be installed separately through the same mechanism, such as \texttt{oneAPI.jl}, \texttt{AMDGPU.jl}, \texttt{Metal.jl}, \texttt{CUDA.jl}, which also downloads the required runtimes and drivers, significantly decreasing the configuration typically required to use accelerators.

\subsection{Reusable GPU Compilation Backends}
\label{sec:reusable_backends}

As Julia can natively inspect and modify Julia source code, as well as the generated LLVM Intermediate Representation (IR) at various stages of compilation, the normal CPU compilation process could be retargeted to different types of IR for GPU accelerators. This functionality is included in the ``\texttt{GPUArrays.jl}'' base package, which afforded a great degree of reusability in compiler infrastructure for the individual backends - for example, the generation of the native PTX instructions on Nvidia platforms, AIR on Apple GPUs and other LLVM IR dialects for AMD and Intel accelerators is achieved in relatively compact individual libraries built on top of \texttt{GPUArrays.jl} - CUDA.jl, Metal.jl, AMDGPU.jl and oneAPI.jl, respectively. For complete details on the unique transpilation architecture, the possibility of writing highly generic and flexible code without sacrificing performance, as well as an implementation of the Rodinia benchmark suite showing that performance is similar to that of the reference platform, the reader is referred to \cite{besard2018effective}.

\subsection{Algorithms Implemented}
\label{sec:algorithms}

The suite of parallel algorithmic building blocks currently included in the first release of AcceleratedKernels.jl is given below, along with some implementation details:

\begin{itemize}
    \item General looping: \texttt{foreachindex} allows the conversion of normal Julia \texttt{for} loops into GPU kernels, with one thread executing each loop iteration. Many pure-Julia functions defined in external packages or the Julia standard libraries can be called from within these kernels, and they will transparently be inlined and transpiled along with the loop body to the target GPU backend.
    \item Sorting elements: \texttt{merge\_sort} and \texttt{merge\_sort\_by\_key} sorts a collection or a pair of keys and payloads kept in separate arrays; both in-place and allocating versions are available.
    \item Sorting indices: \texttt{sortperm} and \texttt{sortperm\_lowmem} compute the array of indices, or index permutation, that would sort a collection; the former algorithm is slightly faster, but requires 50\% more memory than the latter. Again, both in-place and allocating versions are included.
    \item Reduction: \texttt{reduce}, wherein a pairwise operator, or fold, is applied consecutively to all elements in a collection until one final value is computed. As it is executed in parallel with no pre-defined order, no left- or right-associativity can be guaranteed. As for the final result a device-to-host transfer is necessary anyways, a ``\texttt{switch\_below}'' argument is provided which allows the final few intermediate results to be transfered to the host and finish the reduction there, when the cost of kernel launching and device synchronisation are no longer masked by large workloads.
    \item Combined filtering and reduction: \texttt{mapreduce}, similar to \texttt{reduce}, but a unary function is applied to each element before the pairwise operator is used - equivalent to a \texttt{map} followed by a \texttt{reduce}, but without saving the intermediate mapped collection. A great number of algorithms can be implemented on top of \texttt{mapreduce}, such as extracting dimension-wise minima of a set of points (their bounding box), sums, counts, frequencies, etc.
    \item Accumulation: \texttt{accumulate}, or prefix scan, is a common GPU algorithmic building block, where a binary operator is applied such that all elements up to each index are accumulated, or a running total is formed. Both inclusive and exclusive scans are included, with opportunistic look-back \cite{merrill2016single}. Both in-place and allocating versions are included.
    \item Binary search: \texttt{searchsortedfirst} and \texttt{searchsortedlast}, similar to \texttt{std::lower\_bound} and \texttt{std::upper\_bound}, using binary search to find the insertion indices of some elements into a sorted collection such that ordering is maintained. Both in-place and allocating versions are included.
    \item Predicates: \texttt{any} and \texttt{all}, where a unary function returning a boolean is applied to elements in a collection, stopping early once a \texttt{true} is returned (for \texttt{any}), or a \texttt{false} is found (for \texttt{all}). Two algorithms are offered: an optimised one for platforms that allow concurrent writing of competing threads to the same memory location (which is well-defined on modern GPUs if all write the same value - only one thread will do the write, it is just undefined which); on old architectures such as Intel UHD Graphics 620, a conservative algorithm based on \texttt{mapreduce} is included.
\end{itemize}

As GPU VRAM memory is smaller and more expensive than the CPU RAM counterpart, all temporary arrays required by each algorithm are exposed, such that caches in user-code can be reused; all algorithms have been optimised such that all additional memory required is predictably known ahead of time given the input sizes.

The fundamental general parallel looping building block, \texttt{foreachindex} is shown in Algorithm \ref{alg:copy_gpu}, wherein basic Julia loops (equivalent in Algorithm \ref{alg:copy_cpu}) can be converted into parallel code by simply transforming \texttt{for i in eachindex(itr)} into \texttt{AK.foreachindex(itr) do i}. Note that though the \texttt{do-end} block effectively defines a lambda, the objects referenced within \texttt{dst} and \texttt{src} do not have to be explicitly passed into it - instead, they are captured from the surrounding context, with the same performance as explicitly-passed arguments. Benchmarks in the next section show that performance is on par with (and sometimes consistently better than) much older, more mature stacks such as OpenMP.

\begin{algorithm}[H]
\caption{CPU copy kernel implementation as normal Julia code}\label{alg:copy_cpu}
\begin{lstlisting}
function copy_base!(dst, src)
    @assert length(dst) == length(src)
    for i in eachindex(src)
        dst[i] = src[i]
    end
    dst
end
\end{lstlisting}
\label{alg:copy_cpu}
\end{algorithm}

\begin{algorithm}[H]
\caption{GPU and multithreaded CPU copy kernel implementation using AcceleratedKernels.jl}\label{alg:copy_gpu}
\begin{lstlisting}
import AcceleratedKernels as AK

function copy_parallel!(dst, src)
    @assert length(dst) == length(src)
    AK.foreachindex(src) do i
        dst[i] = src[i]
    end
    dst
end
\end{lstlisting}
\label{alg:copy_gpu}
\end{algorithm}

\section{Cross-Architecture Arithmetic Kernels Benchmark}
\label{sec:arithmetic}

GPUs are often used to accelerate relatively simple, numerically-intensive tasks. As extremely simple microbenchmarks are typically difficult to reflect real-world performance, two arithmetic-heavy cases representative of existing algorithms have been chosen here. For brevity, $100,000,000$ 32-bit floating point numbers - which is the most common number type in GPU-accelerated scientific computing - have been tested on all platforms considered. Hardware and software details are given below:

\begin{itemize}
    \item Julia version 1.10.5 is used everywhere applicable.
    \item CPU multithreaded tests all use 10 threads.
    \item On Apple M3 Max (10 performance threads) devices, MacOS Sonoma 14.5, Apple clang version 15.0.0 is used, with LLVM libomp 19.1.0.
    \item On Intel Xeon 8360Y (IceLake architecture, 36 cores, 72 threads) devices, RHEL 8.6 with Linux kernel 4.18.0, GCC 12.3.0 is used with the bundled OpenMP implementation.
    \item On AMD MI210 (gfx90a architecture), same RHEL OS as above, the ROCm 6.1.1 stack is used.
    \item On NVIDIA A100-40 (Ampere architecture), same RHEL OS as above, the CUDA 12.1.1 stack is used.
    \begin{itemize}
        \item Note that both AMD and NVIDIA GPUs are from the same 2022 generation of data centre offerings. 
    \end{itemize}
    \item On NVIDIA L40 (Lovelace architecture), same RHEL OS as above, the CUDA 12.1.1 stack is used.
    \item The Intel GT2 UHD Graphics 620 integrated graphics card is used in a consumer Microsoft Surface 6 laptop with Intel i7-8650U, along with the NEO v24.26.30049+0 Intel Graphics Compute Runtime for oneAPI Level Zero stack.
\end{itemize}

All algorithms in this section have been written so as to be representative of code written by productive, experienced research software engineers, with portability in mind, but without excessive micro-optimisations - to that end, no intrinsics or external packages have been used besides standard libraries. C code has been written following common performance guidelines in a portable C99 subset: data is stored inline, behind pointers; pointer arithmetic is used; standard mathematical functions are used for the correct data type (\texttt{sqrtf}, \texttt{expf}, \texttt{powf}), with no type casts; a single translation unit is compiled; the \texttt{-O2} common optimisation flag has been used, along with \texttt{-Wall -Werror -Wextra} compiler flags, with no warnings produced on any of the compilers and platforms used. Again, to be representative of code written by performance-conscious developers as part of a larger library with common tools - and therefore without micro-optimisations of individual operations, such as manual register placement or swapping mathematical functions with micro-optimised external libraries - Julia code also used the default settings (e.g. \texttt{-O2}) and only tools available in the Base Julia distribution. All code is available in the supplementary materials.

\begin{table*}[htbp]
  \centering
  \caption{Arithmetic benchmark results}
\begin{tabular}{llll|llll}
\multicolumn{4}{l|}{\textbf{Radial Basis Function Kernel}}                                             & \multicolumn{4}{l}{\textbf{Lennard-Jones-Gauss   Potential Kernel}}                                    \\
\textbf{Implementation}                         & \textbf{Device} & \textbf{Arch} & \textbf{Time ($\pm \sigma$) (ms)} & \textbf{Implementation}                         & \textbf{Device} & \textbf{Arch} & \textbf{Time ($\pm \sigma$) (ms)} \\ \hline
\multirow{3}{*}{\textbf{Julia   Base}}          & Apple M3 Max    & aarch64       & 318.35 (2.79)      & \multirow{3}{*}{\textbf{Julia Base}}            & Apple M3 Max    & aarch64       & 219.47 (0.54)      \\
                                                & Intel   8360Y   & x86\_64       & 734.22 (0.29)      &                                                 & Intel 8360Y     & x86\_64       & 335.80 (1.77)      \\
                                                & AMD 7763        & x86\_64       & 799.94 (1.13)      &                                                 & AMD 7763        & x86\_64       & 387.74 (0.25)      \\ \hline
\multirow{3}{*}{\textbf{C}}                     & Apple M3 Max    & aarch64       & 210.57 (1.06)      & \multirow{3}{*}{\textbf{C}}                     & Apple M3 Max    & aarch64       & 1253.0 (4.13)      \\
                                                & Intel   8360Y   & x86\_64       & 641.26 (0.66)      &                                                 & Intel 8360Y     & x86\_64       & 470.61 (1.31)      \\
                                                & AMD 7763        & x86\_64       & 611.23 (0.77)      &                                                 & AMD 7763        & x86\_64       & 501.04 (0.14)      \\ \hline
\multicolumn{4}{l|}{\multirow{3}{*}{}}                                                                 & \multirow{3}{*}{\textbf{C (hand-written powf)}} & Apple M3 Max    & aarch64       & 426.37 (1.24)      \\
\multicolumn{4}{l|}{}                                                                                  &                                                 & Intel   8360Y   & x86\_64       & 381.33 (1.05)      \\
\multicolumn{4}{l|}{}                                                                                  &                                                 & AMD 7763        & x86\_64       & 444.44 (0.13)      \\ \hline
\multirow{3}{*}{\textbf{C   OpenMP}}            & Apple M3 Max    & aarch64       & 23.25 (1.09)       & \multirow{3}{*}{\textbf{C OpenMP}}              & Apple M3 Max    & aarch64       & 28.53 (1.10)       \\
                                                & Intel   8360Y   & x86\_64       & 64.92 (0.05)       &                                                 & Intel 8360Y     & x86\_64       & 53.01 (10.1)       \\
                                                & AMD 7763        & x86\_64       & 61.04 (0.04)       &                                                 & AMD 7763        & x86\_64       & 50.54 (3.95)       \\ \hline
\multirow{3}{*}{\textbf{AcceleratedKernels.jl}} & Apple M3 Max    & aarch64       & 36.33 (0.80)       & \multirow{3}{*}{\textbf{AcceleratedKernels.jl}} & Apple M3 Max    & aarch64       & 27.93 (0.95)       \\
                                                & Intel   8360Y   & x86\_64       & 74.54 (0.05)       &                                                 & Intel 8360Y     & x86\_64       & 49.46 (9.25)       \\
                                                & AMD 7763        & x86\_64       & 82.98 (0.06)       &                                                 & AMD 7763        & x86\_64       & 44.63 (0.03)       \\ \hline
\multirow{5}{*}{\textbf{AcceleratedKernels.jl}} & Apple M3 GPU    &               & 6.24 (0.10)        & \multirow{5}{*}{\textbf{AcceleratedKernels.jl}} & Apple M3 GPU    &               & 10.48 (0.15)       \\
                                                & AMD MI210       & gfx90a        & 2.20 (0.57)        &                                                 & AMD MI210       & gfx90a        & 3.09 (0.33)        \\
                                                & NVIDIA A100-40  & Ampere        & 3.12 (0.00)        &                                                 & NVIDIA A100-40  & Ampere        & 6.03 (0.00)        \\
                                                & NVIDIA L40      & Lovelace      & 2.88 (0.03)        &                                                 & NVIDIA L40      & Lovelace      & 5.39 (0.06)        \\
                                                & Intel GT2 UHD   & CometLake     & 100.68 (1.99)      &                                                 & Intel GT2 UHD   & CometLake     & 221.68 (5.39)     
\end{tabular}
\label{tab:arithmetic}%
\end{table*}%

\subsection{Radial Basis Function Kernel}
\label{sec:rbf}

A radial-basis function-like kernel - as used in support vector machines, Gaussian kernels, and neural network activation functions - is tested here, with relatively heavy numerical operations (exponentiation, division and a square root), but with few steps and no branching is given in Algorithm \ref{alg:rbf}. 100 million 3D points are considered, with the X, Y, Z coordinates stored inline (same storage in both Julia and C).

\begin{algorithm}[H]
\caption{Example AcceleratedKernels.jl implementation of the Radial Basis Function arithmetic benchmark}\label{alg:rbf}
\begin{lstlisting}
AK.foreachindex(rbf) do i
    rbf[i] = exp(-1/(1-sqrt(v[1,i]^2 +
                            v[2,i]^2 +
                            v[3,i]^2)))
end
\end{lstlisting}
\label{alg:rbf}
\end{algorithm}

As shown in Table \ref{tab:arithmetic}, for such simple arithmetic kernels ARM-based architectures (aarch64) consistently perform better in the CPU space, for all implementations. C code is 33.9\% faster in the single-threaded case than the Julia code for ARM, and 12.7\% faster on x86\_64. For the multithreaded case, OpenMP achieves 89.6\% strong scaling on M3 Max and 98.8\% strong scaling on x86\_64, possibly due to the greater stack maturity on the latter; AcceleratedKernels.jl produces similar figures with 87.6\% and 98.5\% strong scaling on ARM and x86\_64. Therefore, the performance of Julia base threads is equivalent to that of the much older OpenMP, even in the most straightforward to optimise OpenMP case - while Julia threads afford much greater flexibility than a directives-based approach. C code seems to produce fewer instructions than Julia, though Julia does insert floating-point correctness checks for \texttt{sqrt} which are propagated via exceptions, which are heavier computationally. GPU codes, depicted in the last rows of Table \ref{tab:arithmetic} all provide consistent, good speed-ups over the CPU counterparts; interestingly, though both AMD and NVIDIA GPUs are from the same generation - and in spite of the NVIDIA stack being more mature and widely used - AMD MI210 is 29.5\% faster than the NVIDIA A100-40.

Note that both C and Julia implementations used the power operator for squaring terms - and all C compilers and Julia replaced it with multiplication (i.e. transforming \texttt{x\^{}2} into \texttt{x*x}) in the compiled binaries.

\subsection{Lennard-Jones-Gauss Potential Kernel}
\label{sec:ljg}

A more complex arithmetic benchmark is considered here as the Lennard-Jones-Gauss (LJG) potential (Algorithm \ref{alg:ljg}) used in molecular dynamics (MD) to model more complex assemblies such as polymers and colloidal systems, including a cutoff distance beyond which forces are considered small enough to be neglected \cite{zhou2018thermodynamic}. More terms and calculation steps are used, with stronger dependencies between them; importantly, there is a difficult to predict branching \texttt{if} statement, which results in operation serialisation on GPUs, as in-sync groups of threads (warps in NVIDIA nomenclature, wavefronts for AMD) would have to wait for each branch in turn; CPU cores can typically execute entirely different instructions without waiting. Note that while in MD simulations interactions are computed pairwise between all atoms that are closer than a cutoff distance - further accelerated with geometrical data structures such as neighbour lists \cite{thompson2022lammps} - in order to measure arithmetic performance two separates arrays of atomic positions are considered (100 million atoms with their X, Y, Z coordinates stored inline; exactly the same storage used in both Julia and C). The constants used are \texttt{epsilon=1, sigma=1, r0=1.5, cutoff=3}, and they are passed into the function at runtime so that constant propagation cannot optimise them out.

\begin{algorithm}[H]
\caption{Example AcceleratedKernels.jl implementation of the Lennard-Jones-Gauss potential arithmetic benchmark}\label{alg:ljg}
\begin{lstlisting}
AK.foreachindex(energy) do i
    r = sqrt(
        (atoms1[1,i]-atoms2[1,i])^2 +
        (atoms1[2,i]-atoms2[2,i])^2 +
        (atoms1[3,i]-atoms2[3,i])^2)
    if r < cutoff
        lj_energy = 4*epsilon*(
            (sigma/r)^12-(sigma/r)^6)
        g_energy = A*exp(-(r-r0)^2 /
            (2*sigma^2))
        energy[i] = lj_energy+g_energy
    else
        energy[i] = 0
    end
end
\end{lstlisting}
\label{alg:ljg}
\end{algorithm}

The first unexpected result shown in Table \ref{tab:arithmetic} is the massive times recorded for the base C implementation (5.7 times slower than Julia on ARM, and 1.4 times slower on x86\_64), as well the resulting speed-ups with OpenMP parallelisation - 43.9 times improvement on ARM when using only 10 threads. Upon inspecting the disassembled shared libraries produced on the ARM (Clang) and x86\_64 (GCC) compilers, both showed that the integer powers used in e.g. \texttt{powf(sigma/r, 6)} were not optimised to simple multiplication, and instead showed 10 calls to the C standard math library \texttt{powf} function, which iteratively numerically computes powers; however, when using the OpenMP directive, only 2 \texttt{powf} calls are emitted. Indeed, running the OpenMP version even with a single thread - though the kernel is written exactly the same - results in a $>4$-fold improvement over the normal C compilation. It seems \texttt{powf} is much slower on ARM than on x86\_64. To validate this, another C kernel was written where the \texttt{powf} calls have manually been substituted with multiplication (i.e. \texttt{pow3=x*x*x;pow6=pow3*pow3;pow12=pow6*pow6}), thus showing a 2.94-fold improvement on ARM and 1.23 on x86\_64; still, the OpenMP version (which does not have the hand-written exponentiation) times are much faster than as would result from multithreading, showing the equivalent of 149.4\% strong scaling on ARM. While it could be argued that Apple ARM CPUs are more recent architectures with less mature optimisations available in modern compilers, it seems that Julia does not suffer from these issues, even though both Clang and the Julia compiler are built on LLVM. AcceleratedKernels.jl multithreading shows 78.6\% strong scaling on ARM and 67.9\% on x86\_64; relative to the hand-written exponentiation, C OpenMP results in 71.9\% strong scaling on x86\_64. Thus, interestingly, Julia proves to be more consistent with the performance of numerical code than C in the benchmarks considered here. Julia converted all integer powers to multiplication; finally, in all CPU cases Julia consistently produced better performing code than GCC and Clang. The GPU results follow the same trends as for the Radial Basis Function benchmark, with the AMD MI210 being 1.95-times faster than the NVIDIA A100-40; still, the Apple GPU (on a consumer laptop) shows hopeful results, being within a factor of 1.74 of the A100-40, a data centre-grade GPU. While most CPU results were faster on the LJG potential benchmark, the GPU timings are all slower - possibly due to the branching nature of the kernel implemented, which is not as performant on GPUs in general.

\section{Cross-Device Sorting Benchmark in MPISort.jl}
\label{sec:sorting_benchmark}

In order to benchmark the large-scale scalability of AcceleratedKernels.jl algorithms, a series of tests have been run in June 2024 over the 208 NVIDIA A100 GPUs of the Baskerville Tier 2 UK HPC cluster. These tests highlighted a few key points:

\begin{itemize}
    \item The exceptional composability of Julia libraries, which transparently used Julia Base CPU sorters, AcceleratedKernels.jl sorting kernels, and NVIDIA Thrust C++ sorting algorithms together with the MPISort.jl multi-node sorter, as well as the NVLink high-speed GPU-to-GPU direct interconnects available on Baskerville which can be used through the hardware-optimised MPI library. Such specialised MPI implementations can again be transparently used via the MPI.jl Julia package \cite{byrne2021mpi}.
    \item World-class throughput being possible to be achieved with user-friendly Julia algorithms on Tier 2 HPC clusters as with esoteric C++ sorters on world-leading supercomputers: the highest sorting throughput reported in literature is the 900 GB/s achieved on the 262,144 AMD cores of the CRAY XK7 "Titan" platform at the Oak Ridge National Laboratory \cite{sundar2013hyksort}; in comparison, we reached 850 GB/s on 208 NVIDIA A100 GPUs with NVLink interconnects. 
    \item Good performance and scaling being achieved with sorters written in the AcceleratedKernels.jl backend-agnostic, unified, transpiled codebase as with highly-optimised NVIDIA Thrust algorithms.
\end{itemize}

The NVIDIA Thrust library of parallel algorithm building blocks - recently merged into the CUDA Core Compute Libraries (CCCL) - has been exposed to Julia via a C Foreign Function Interface (FFI). CUDA arrays are allocated in Julia using the CUDA.jl package; while their memory is managed by the Julia garbage collector, the pointer to the internal CUDA memory can be passed via C FFI. Templated C++ algorithms were written converting a given raw pointer to the Thrust \texttt{thrust::device\_ptr} wrappers, which are then used to call the templated Thrust algorithms. The templates have been instantiated into explicitly-defined C functions annotated with the \texttt{extern "C"} specifier, which were then compiled into a shared library; for this benchmark, only numerical types were explicitly defined, namely \texttt{int16\_t, int32\_t, int64\_t, \_\_int128, float, double}. Julia offers native C-calling capabilities from symbols defined in shared libraries with no additional steps.

\subsection{MPISort Algorithm}
\label{sec:mpisort}

The MPISort.jl library, developed by the authors, implements the ``Sampling with Interpolated Histograms Sort'' - or SIHSort - algorithm for multi-node sorting. It is based on the sample sort algorithm, using MPI communication to find ``splitters'' between MPI ranks such that elements between splitter N and splitter N + 1 will end up on rank N \cite{frazer1970samplesort}. It requires the use of two rank-local sorting steps, where the initial data is sorted, then after the final rearrangement across ranks following the splitters. While it works for any comparison-based data, additional optimisations were made for numerical elements, again being generic by virtue of Julia's type system. Significant optimisations were made to reduce MPI communication, for example having counters hidden at the end of integer arrays, merging their functionality, such that the number of MPI calls is minimised; to the best of our knowledge, among non-IO based algorithms, this implementation uses the least amount of MPI communication. Except for the final redistribution of data following the splitters, the memory footprint only depends on the number of ranks involved, hence improving its scalability. The library is a registered open-source Julia package, available at https://github.com/anicusan/MPISort.jl.

\subsection{Baskerville HPC Architecture}
\label{sec:baskerville}

The Baskerville Tier 2 high-performance computing (HPC) cluster at the University of Birmingham, UK (link: www.baskerville.ac.uk/) comprises 52 SD650-N V2 liquid cooled compute trays, each with 2 Intel Xeon 8360Y 36-core CPUs, 512GB RAM, 4 NVIDIA A100 GPUs, NVIDIA HGX-100 GPU planar, and NVIDIA Mellanox Infiniband - along with Lenovo Neptune direct liquid cooling. A key highlight of its architecture is that all GPUs are meshed with NVLink high-speed direct GPU-to-GPU interconnects, such that data can be transferred between GPUs directly, without incurring device-to-host copies; as shown later in Figure \ref{fig:cost}, this is crucial in making data centre GPUs in high-performance computing applications more cost-effective than their CPU counterparts.

\subsection{Benchmarks}
\label{sec:mpisort_benchmarks}

In all algorithms below, ``CPU Transfer'' indicates that MPI communication happens over CPU RAM - for GPU algorithms, that implies a device-to-host data transfer; ``NVLink Transfer'' means that direct GPU-to-GPU MPI communication is used over NVLink interconnects. An MPI ``rank'' is used here to refer to a CPU core - for the Julia Base single-threaded CPU sorting algorithm - or a GPU device for the other algorithms.

\begin{figure}[htbp!]
\centering
\includegraphics[width=0.99\linewidth]{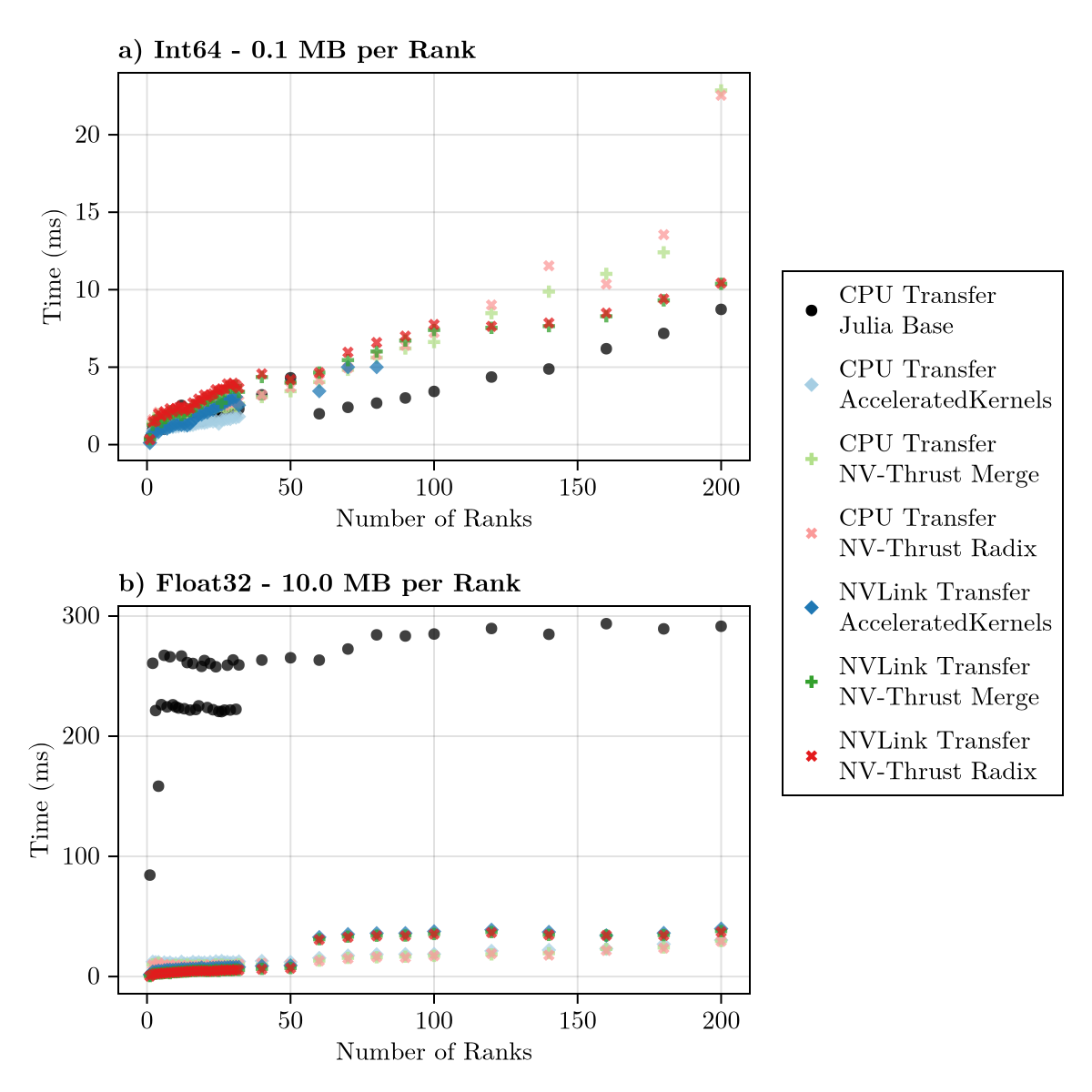}
\caption{Weak scaling tests for the CPU and GPU sorting algorithms at low data sizes per rank. Nomenclature: ``CC-JB'' is the Julia Base algorithm with CPU-CPU MPI communication; the ``GC'' prefix qualifies MPI communication over CPU RAM, incurring a device-to-host copy; the ``GG'' prefix represents direct GPU-to-GPU communication over NVLink interconnects; the ``AK'' suffix stands for the AcceleratedKernels.jl merge sort algorithm; ``TM'' is the NVIDIA Thrust merge sort; ``TR'' is the NVIDIA Thrust radix sort.}
\label{fig:low_times}
\end{figure}

As shown in panel $a)$ Figure \ref{fig:low_times}, for very low data sizes per MPI rank (CPU or GPU) - e.g. 0.1 MB per rank, corresponding to 25,000 Int32 values - the CPU algorithms consistently outperform the GPU ones. As expected, at greater data sizes (panel $b)$, 10 MB per rank, or 2,500,000 Int32 values) GPU algorithms can be an order of magnitude faster, as their massive parallelism - at the expense of higher data transfer costs - is better exploited. Therefore, in the next figures only GPU algorithms will be depicted at higher data sizes for better assessment.

\begin{figure*}[htbp!]
\centering
\includegraphics[width=0.85\linewidth]{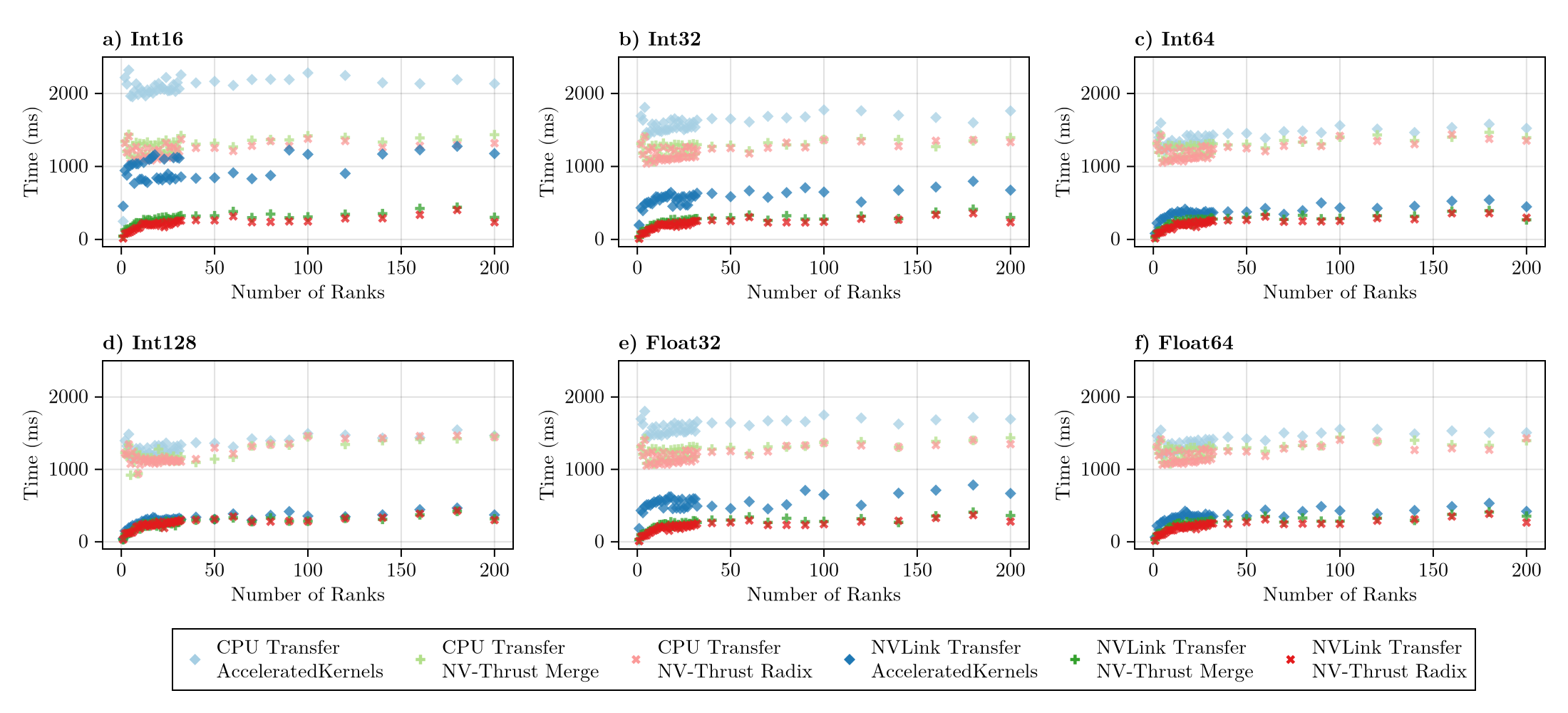}
\caption{Weak scaling of the GPU sorting algorithms for the data types considered at 1 GB of data per rank.}
\label{fig:high_times}
\end{figure*}

As shown in the weak scaling tests in Figure \ref{fig:high_times}, algorithms using direct GPU-to-GPU interconnects (darker hues) are consistently, significantly outperforming the other ones. A positive result is that once communication becomes the main performance factor (above 12 GPUs, corresponding to three nodes), the weak scaling of the MPISort algorithm remains relatively flat across all local sorters used, thus showing very good scaling with bigger problem sizes. For smaller data types such as Int16, the special-cased optimisations for numbers included in the NVIDIA Thrust library become more important, and thus faster than AcceleratedKernels.jl; for example, radix sort iterates over each individual bit of the numerical data type to be sorted. For larger data types, this difference becomes smaller, such that the Int64, Int128, and Float64 cases produce comparable timings between the AcceleratedKernels.jl local sorters and Thrust ones - indeed for the Int128 case becoming all but indistinguishable.

% \begin{figure*}[htbp!]
% \centering
% \includegraphics[width=0.9\linewidth]{resources/weak_scaling.png}
% \caption{Weak scaling efficiency for the GPU sorting algorithms for the data types considered.}
% \label{fig:weak_efficiency}
% \end{figure*}

\begin{figure*}[htbp!]
\centering
\includegraphics[width=0.85\linewidth]{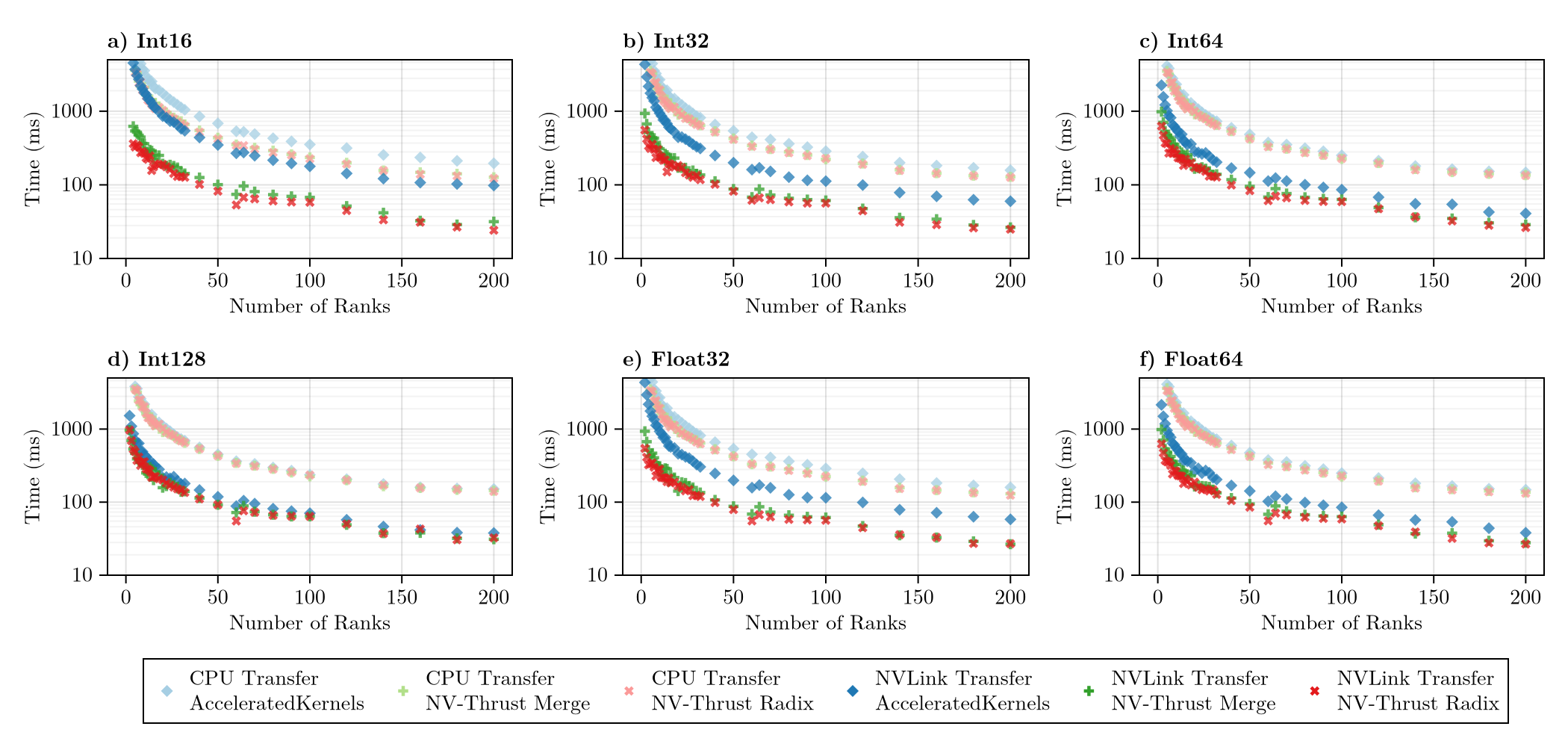}
\caption{Strong scaling of the GPU sorting algorithms for the data types considered at 16 GB of data divided over the ranks.}
\label{fig:strong_scaling}
\end{figure*}

The strong scaling tests shown in Figure \ref{fig:strong_scaling} again depict a very strong difference between the algorithms using the NVLink direct GPU-to-GPU interconnects (in darker hues) and the ones that do not, becoming more significant as more ranks are used. A positive result here is that all algorithms show relatively good strong scaling, seemingly with some improvement still to be had even beyond the 200 GPUs tested here - though, as expected, showing diminishing returns.

\begin{figure*}[htbp!]
\centering
\includegraphics[width=0.85\linewidth]{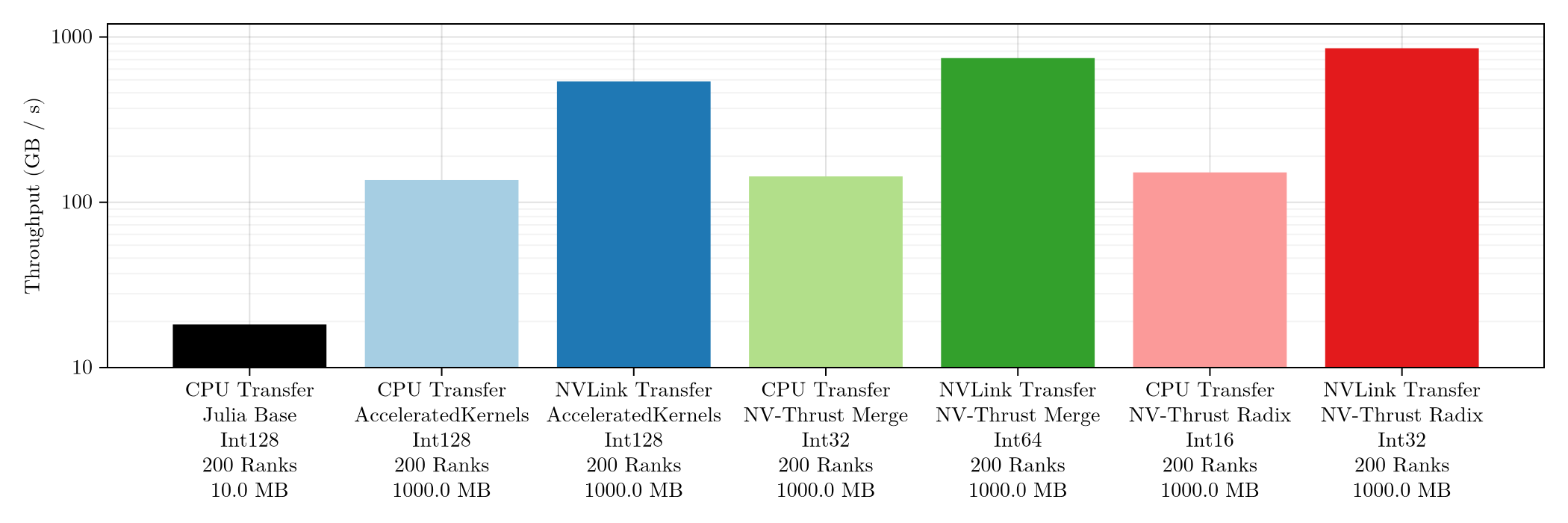}
\caption{Maximum throughput achieved for the CPU and GPU sorting algorithms, including the test case data type and size per rank for which each best was recorded.}
\label{fig:throughput}
\end{figure*}

Among all tests conducted, the maximum throughput achieved (GB of data sorted per second) has been recorded, along with the test case for which it was found; these results are shown in Figure \ref{fig:throughput}. Again, there is a stark difference between the algorithms using GPU-to-GPU interconnects (the ones prefixed with ``GG'') than the ones that do not (prefixed with ``GC''). Still, even with an additional device-to-host copy, the slowest GPU algorithm is 7.48 times faster than the equivalent CPU algorithm (depicted in black), with not much differentiation between ``GC-*'' algorithms. The three fastest throughputs achieved, for the NVIDIA Thrust radix sort (855 GB / s), Thrust merge sort (745 GB / s) and AcceleratedKernels.jl (538 GB / s), are all over an order of magnitude faster than the CPU algorithm, and on average 4.93 times faster than the algorithms not using NVLink interconnects. Another noteworthy finding is that the CPU and AcceleratedKernels.jl were fastest for larger, more complex data types (Int128), while the Thrust algorithms were faster for smaller data types; all maxima were found when sorting signed integers.

\begin{figure}[htbp!]
\centering
\includegraphics[width=0.85\linewidth]{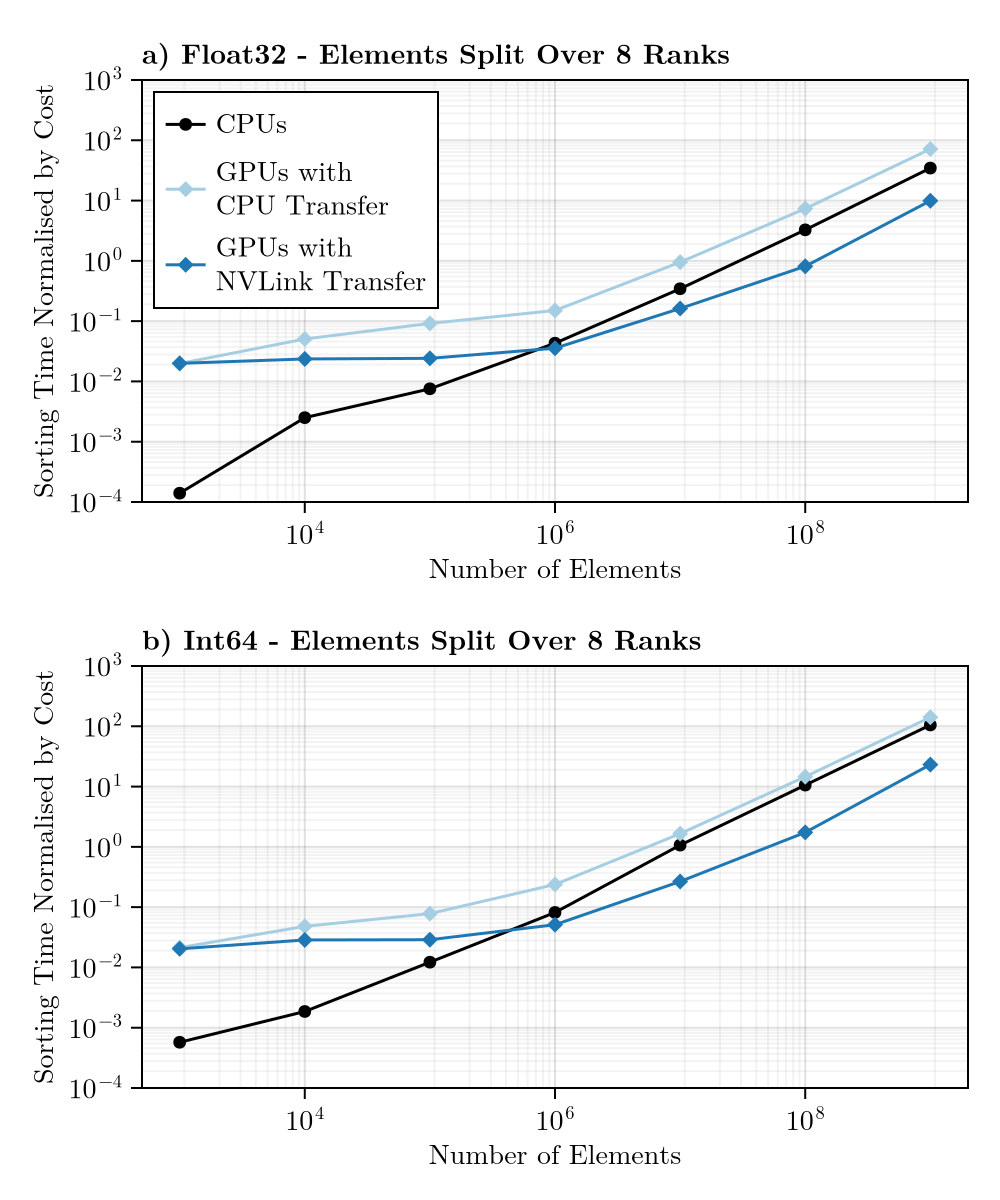}
\caption{Sorting times normalised by a 22 GPU-to-CPU combined capital, running and environment cost ratio.}
\label{fig:cost}
\end{figure}

An important consideration in the development of GPU-centric HPC clusters is their cost - in absolute terms, GPUs are more expensive (capital cost), use more power (running cost) and produce more CO$_2$ emissions (environmental cost) than CPUs. In order to compare their improved performance over the higher costs, the GPU sorting times were normalised by a factor of 22, representing the combined excess costs over the lifetime of a typical GPU-centric HPC; while a rough figure, the University of Birmingham Advanced Research Computing team, which is in charge of developing both the BlueBEAR (CPU-centric) and Baskerville (GPU-centric) Tier 2 HPC resources, have validated this number. As shown in Figure \ref{fig:cost}, when sorting over one million elements, for both the Float32 and Int64 cases, the additional costs of GPUs over CPUs become economically justifiable in communication-heavy HPC tasks (a prime example of which being multi-node data sorting) only when using direct GPU-to-GPU interconnects.

\section{Conclusion}

The AcceleratedKernels.jl library introduced in this paper showed that code flexibility, programmer productivity and high performance can be achieved altogether using the unique architecture of a transpilation-based unified codebase of parallel algorithms. As detailed in Section \ref{sec:alternatives}, among cross-architecture programming models, this approach provides the best hardware support after Vulkan (being the only two natively targeting Nvidia, AMD, Intel and Apple accelerators), while the former provides the lowest implementation and usage complexities. As algorithms written in the KernelAbstractions.jl Julia-based kernel language are transpiled into the native intermediate representation (IR) of each target platform (PTX for NVIDIA, AIR for Apple, other LLVM IR dialects for Intel and AMD), we benefit from all high-level optimisations available in Julia, as well as all optimisations included in the official, native software toolchains. A highlight of its ease of use described in Section \ref{sec:algorithms} is the possibility of converting most normal Julia \texttt{for} loops into parallel code (both statically-partitioned on CPU threads, or one iteration-per-thread on GPUs) by simply substituting the \texttt{for i in eachindex(itr)} construct with \texttt{AK.foreachindex(itr) do i}; importantly, many functions defined in external packages or the Julia Base standard library can be called from within the kernel bodies with no special-casing, which are then inlined and compiled along with the rest of the kernel on the target platform. As shown in the arithmetic benchmarks (Section \ref{sec:arithmetic}), Julia performance is on par, and sometimes exceeding that of performance-conscious, portable C and OpenMP code; surprisingly, performance in numerical code can be more consistent and predictable in Julia than C; very good speed-ups were seen across the Apple GPU, AMD MI210 and NVIDIA A100-40 accelerators tested. The excellent composability of Julia code has been shown in Section \ref{sec:mpisort}, wherein Julia Base CPU sorters, AcceleratedKernels.jl GPU merge sorters, and NVIDIA Thrust C++ merge and radix sorters were coupled with a multi-node MPISort algorithm, transparently making use of hardware-specialised MPI implementations using the Baskerville Tier 2 HPC NVLink direct GPU-to-GPU interconnects - all without special-casing any of the libraries. Very good weak scaling has been seen across all algorithms, with indistinguishable performance for larger, more complex data types between AcceleratedKernels.jl Julia sorters and Thrust C++ sorters, but with more prominent differences where small numerical data types were special-cased in Thrust. World-class 538-855 GB/s sorting throughputs were achieved on 200 GPUs, comparable with the highest reported figure of 900 GB/s achieved on 262,144 CPU cores. Finally, using direct GPU-to-GPU NVLink interconnects were shown to consistently provide significant speed-ups, being on average 4.93 times faster than cases not using them; normalising the sorting performance of GPU algorithms by a combined capital, running and environmental cost resulted in communication-heavy HPC workloads only becoming economically viable if direct GPU-to-GPU interconnects are used.

\section*{Supplementary Materials}
The first release of AcceleratedKernels.jl as used in this paper has been archived for reproducibility purposes on Zenodo (DOI: 10.5281/zenodo.13840912). The benchmarking code, HPC runtime logs, disassembled shared libraries, analysis and plot-making scripts have been archived separately (DOI: 10.5281/zenodo.13840910). The MPISort.jl library is similarly archived (DOI: 10.5281/zenodo.13840921).

\section*{Acknowledgements}
The computations described in this paper were performed using the University of Birmingham's BlueBEAR and Baskerville HPC services (link: www.baskerville.ac.uk/), which provide a High Performance Computing service to the University's research community. See http://www.birmingham.ac.uk/bear for more details.

% References
\printbibliography % Output the bibliography

\section{Biography Section}
 

\begin{IEEEbiography}[{\includegraphics[width=1in,height=1.25in,clip,keepaspectratio]{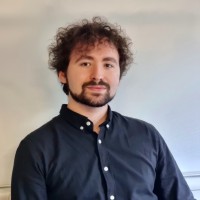}}]{Andrei-Leonard Nicușan}
is a final-year doctoral researcher in the
University of Birmingham’s School of Chemical Engineering and CTO of EvoPhase Ltd., an AI in industry spinout. He published featured articles and Scientific Highlights on machine learning-based algorithms, metaprogramming-driven evolutionary optimisation, simulational-experimental calibration and positron imaging, based on which he won the 2024 IChemE Young Engineers Award for Innovation and Sustainability; his open-source frameworks are actively being used in academia and industry, with work in partnership with GlaxoSmithKline winning the 2023 “Best Use of HPC in Industry” award from HPCWire.
\end{IEEEbiography}

\begin{IEEEbiography}
[{\includegraphics[width=1in,height=1.25in,clip,keepaspectratio]{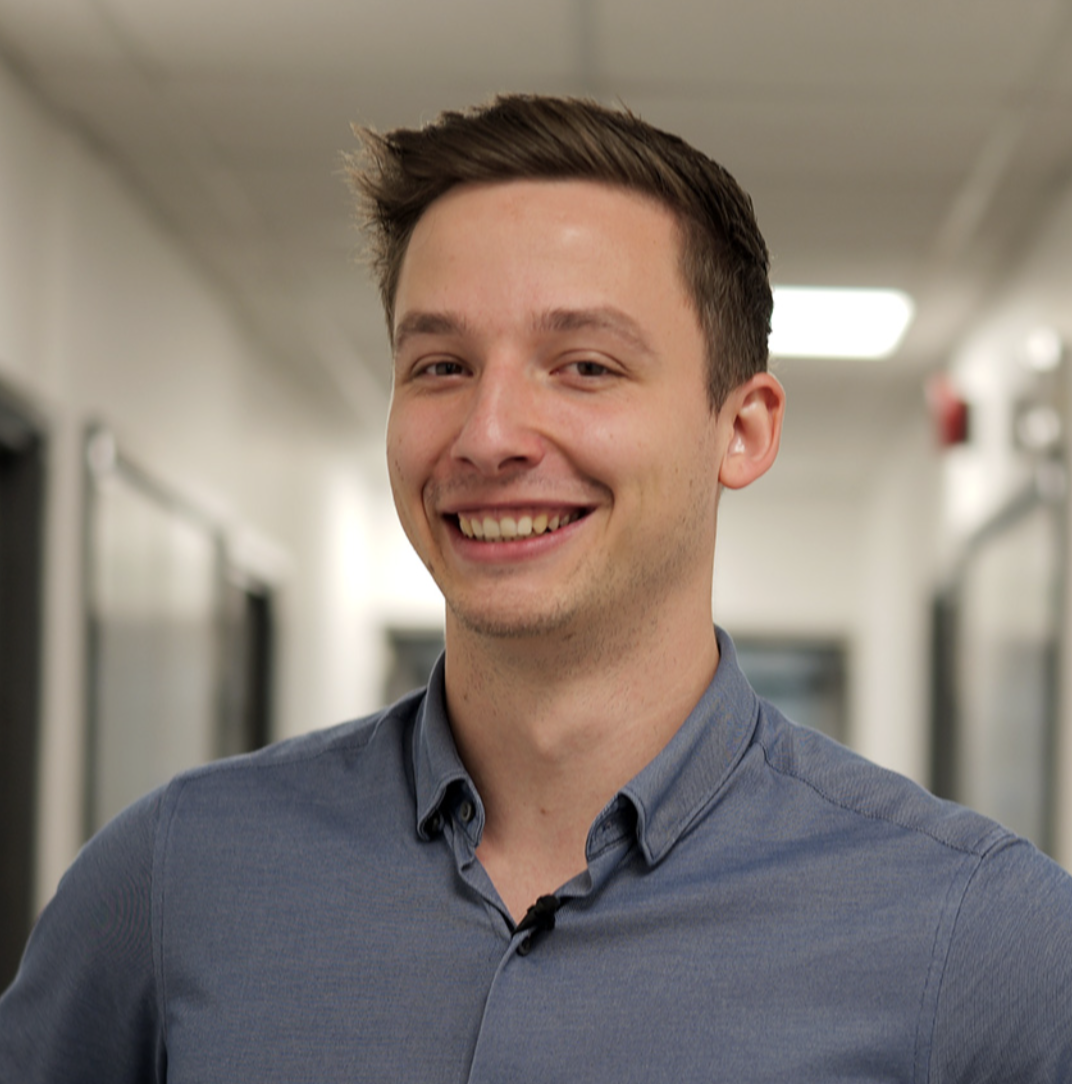}}]{Dominik Werner}
is a final-year doctoral researcher in the University of Birmingham's School of Chemical Engineering and CEO of EvoPhase Ltd. He has experience in imaging and simulating fluidised beds, zero-gravity granular dampers and self-optimising granular systems. His digital models of industrial-scale fluidised beds, high-shear mixers, conveying equipment and powder characterisation instruments are actively used by companies such as Recycling Technologies, FMC, JDE, Mondelez, P\&G, Unilever and AstraZeneca.
\end{IEEEbiography}

\begin{IEEEbiography}
[{\includegraphics[width=1in,height=1.25in,clip,keepaspectratio]{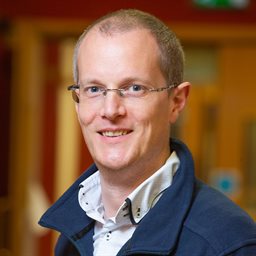}}]{Simon Branford}
is the Deputy Leader of the Research Software Group and Principal Research Software Engineer of the Advanced Research Computing group at the University of Birmingham, with experience in hybrid Monte Carlo algorithms for linear algebra problems. He is an EasyBuild maintainer and a member of the UKRI Tier-2 high performance computing Technical Working Group.
\end{IEEEbiography}

\begin{IEEEbiography}
[{\includegraphics[width=1in,height=1.25in,clip,keepaspectratio]{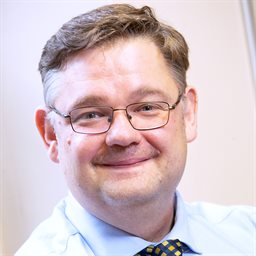}}]{Simon Hartley}
is a Senior Research Software Engineer of the Advanced Research Computing group at the University of Birmingham, with experience in software for dielectric materials, MRI analysis and Genetic Epidemiology, Computer Vision Systems for Monitoring Civil Engineering, and has built autonomous robots which have been deployed in nuclear power stations.
\end{IEEEbiography}

\begin{IEEEbiography}
[{\includegraphics[width=1in,height=1.25in,clip,keepaspectratio]{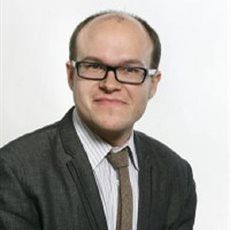}}]{Andrew J. Morris}
is Professor of Computational
Physics at the University of Birmingham (UoB). He is Director of the
UKRI Tier 2 Baskerville GPU Accelerated computer housed at UoB and
Co-investigator of the UKRI Tier 2 Sulis high-throughput computer run
by the HPC Midlands+ consortium. Within UoB he chairs the Research
Computing Management committee and leads the materials simulation
and modelling discussion group. He is lead author of the OpenSource
OptaDOS code, aiding fundamental characterisation of materials at over 16 universities and national facilities within the UK, US, Canada, Japan, China and Israel.
\end{IEEEbiography}

\begin{IEEEbiography}
[{\includegraphics[width=1in,height=1.25in,clip,keepaspectratio]{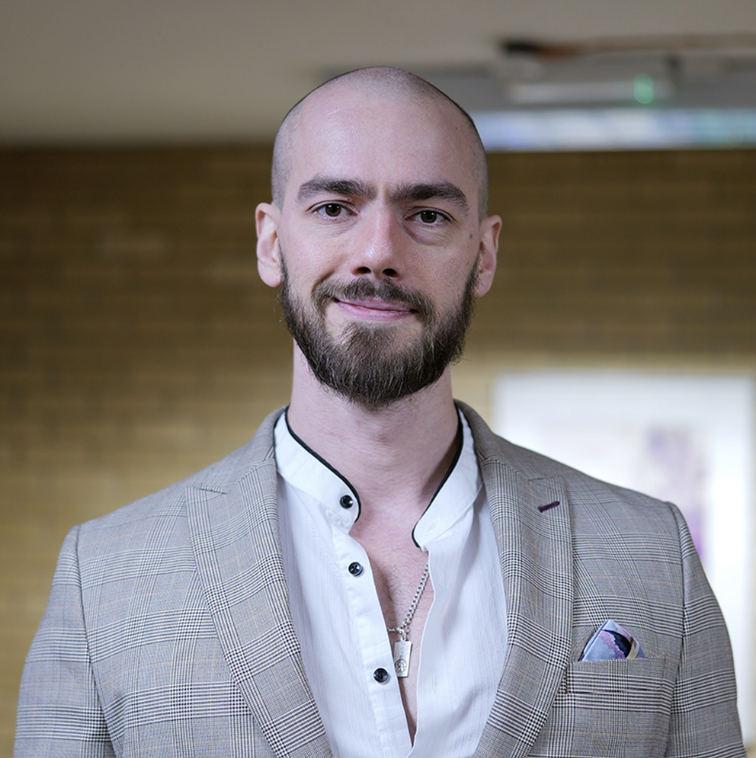}}]{Kit Windows-Yule}
is a Turing Fellow, Royal Society Industry Fellow,
a two-time Royal Academy of Engineering Fellow, an Innovate UK
BridgeAI Independent Scientific Advisor, Associate Professor
of Chemical Engineering at the University of Birmingham (UoB), and CSO of EvoPhase Ltd. He is also Deputy Chair of UoB’s Research Computing Management Committee. He is leading projects in developing novel plastic recycling methods, novel methods of blood-flow imaging for the diagnosis of cardiovascular disease, and diverse industry-funded work in the pharmaceutical, food, agriculture, chemical, personal care and green energy sectors with companies including AstraZeneca, GlaxoSmithKline, Mondelez, Johnson Matthey, Unilever and the French Petroleum Institute’s Energies Nouvelles arm.
\end{IEEEbiography}

\end{document}